\documentclass[10pt]{article}

\usepackage{amsmath}
\usepackage{amssymb}

\usepackage{graphicx}
\usepackage{color}

\usepackage{cite}

\usepackage{color} 

\usepackage[labelfont=bf,labelsep=period,justification=raggedright]{caption}

\makeatletter
\renewcommand{\@biblabel}[1]{\quad#1.}
\makeatother

\date{}

\begin{document}

\begin{flushleft}
{\Large
\textbf{A Positive Feedback at the Cellular Level Promotes Robustness and Modulation at the Circuit Level}
}
\\
Julie Dethier$^{1,2,*}$, 
Guillaume Drion$^{1,3,4*}$, 
Alessio Franci$^{1,5}$,
Rodolphe Sepulchre$^{1,5}$
\\
\bf{1} Department of Electrical Engineering and Computer Science, University of Li\`ege, Li\`ege B-4000, Belgium
\\
\bf{2} Department of Mechanical and Aerospace Engineering, Princeton University, Princeton, NJ 08544, USA
\\
\bf{3} Laboratory of Pharmacology and GIGA Neurosciences, University of Li\`ege, Li\`ege B-4000, Belgium
\\
\bf{4} Volen Center for Complex Systems, Brandeis University, Waltham, MA 02454, USA
\\
\bf{5} Department of Engineering, University of Cambridge, Cambridge CB2 1PZ, UK
\\
\bf{*} These authors contributed equally to this work
\end{flushleft}

\section*{Abstract}

The paper highlights the role of a positive feedback gating mechanism at the cellular level in the robustness and modulation properties of rhythmic activities at the circuit level. The results are presented in the context of half-center oscillators, which are simple rhythmic circuits composed of two reciprocally connected inhibitory neuronal populations. Specifically, we focus on rhythms that rely on a particular excitability property, the post-inhibitory rebound, an intrinsic cellular property that elicits transient membrane depolarization when released from hyperpolarization. Two distinct ionic currents can evoke this transient depolarization: a hyperpolarization-activated cation current and a low-threshold T-type calcium current. The presence of a slow activation is specific to the T-type calcium current and provides a slow-positive feedback at the cellular level that is absent in the cation current. We show that this slow-positive feedback is necessary and sufficient to endow the network rhythm with physiological modulation and robustness properties. This study thereby identifies an essential cellular property to be retained at the network level in modeling network robustness and modulation.

\section*{Introduction}

Biological rhythms play a major role in the functioning of the brain but much remains to be understood regarding their control, regulation, and function. Many advances in this important question have come from experimental and computational studies of central pattern generators (CPGs), which endogenously produce precise rhythmic outputs directly related to motor functions~\cite{Grillner1985, Getting1989, Marder1996, Marder2001, Harris-Warrick2011}. In this effort, experimental work benefits from computational models but models at the circuit level usually rely on mathematical simplifications at the component level. The question of which cellular details must be retained at the network level is largely open ~\cite{Marder2006}.

Motivated by this general question, we highlight a simple feedback mechanism at the cellular level that has a key influence on circuit robustness and modulation. We illustrate this property via the computational study of an archetype model of CPG circuits, the \emph{half-center oscillator} (HCO): two neuronal populations that do not oscillate in isolation, but oscillate in an antiphase rhythm when reciprocally connected~\cite{Brown1911, Perkel1974, Satterlie1985}. Because of the widespread occurrence of this circuit motif, the mechanisms have been extensively studied, both computationally and experimentally~\cite{Calabrese1992, Marder1996, Harris-Warrick2011, Skinner1993, Skinner1994, Wang1992, Daun2009}. A specific cellular excitability property, the \emph{post-inhibitory rebound} (PIR)~\cite{Perkel1974}, and its two specific ionic currents, $I_H$ and $I_{Ca,T}$~\cite{Angstadt2005}, are well-known key players in circuit oscillations.

Previous studies have focused on distinguishing those two currents from their contribution to the escape or release mechanism~\cite{Wang1992, Skinner1994, Daun2009} and their modulation of rhythmic activity~\cite{Cymbalyuk2002, Sorensen2004, Olypher2006, Doloc-Mihu2011}. This paper highlights that those two currents differ in another simple yet fundamental aspect: : both generate a PIR, but only one of them acts, through its slow activation, as a source of positive feedback in the network rhythm timescale. The present paper demonstrates through a computational study that this particular feedback is fundamental for the robustness and modulation properties of the circuit rhythm, and that its absence is detrimental both to robustness and modulation at the circuit level.

Our results predict that PIR \textit{per se} is not a sufficient cellular excitability property to be retained at the network level. In addition, the current \emph{slow-regenerative} regenerative (slow-positive-feedback) nature must be retained as an important dynamical parameter. We emphasized in previous work the importance of this positive feedback at the cellular level in defining bursting excitability~\cite{Franci2014} and its widespread regulation in different neuron types~\cite{Franci2013}. The present paper prolongs this work, moving from the role of regenerativity at the cellular level to its importance at the circuit level.

\section*{Results}

\subsection*{Slow activation of T-type calcium channels is critical to robustness of network rhythmic activity}

To assess the role of cellular properties in network rhythms, we consider one of the simplest and most studied networks: the half-center oscillator (HCO). The network rhythm results from the mutual inhibition (I) of two neurons that do not oscillate endogenously in isolation~\cite{Brown1911, Perkel1974, Satterlie1985}. HCOs have been identified at the core of most endogenous rhythmic circuits, such as CPGs governing locomotion ~\cite{Brown1911, Perkel1974, Satterlie1985, Calabrese1992, Daun2009} or respiration~\cite{Grillner1985, Butera1999a, Daun2009}. Oscillations in HCOs are triggered by an external pulse of hyperpolarizing current. When released from hyperpolarization, the cell generates a burst-like transient depolarization with one or more spikes. This activity hyperpolarizes the other cell via the inhibitory synaptic connection, which in turn triggers a transient burst. The cycle repeats leading to an antiphase rhythm between the two neurons.

\begin{figure}[!h]
\begin{center}
\includegraphics[width=0.85\textwidth]{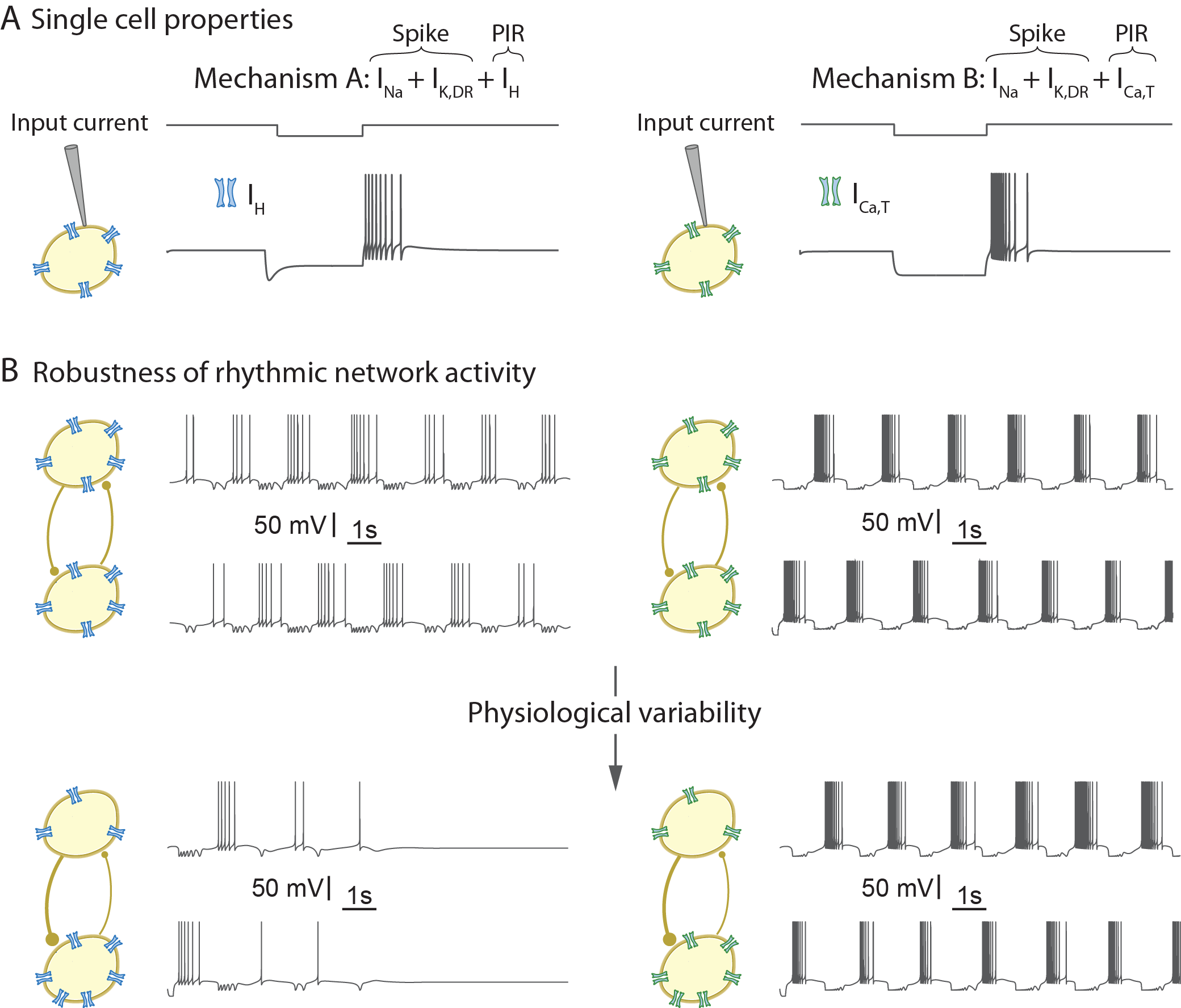}
\end{center}
\caption{
{\bf Network rhythmic activities generated by distinct post-inhibitory-rebound mechanisms strongly differ in their robustness properties.} A:~Mechanism A generates a PIR with a $I_H$-type current and Mechanism B generates a PIR with a slowly activating $I_{Ca,T}$-type current (see Methods for cellular models). ). B:~In a half-center network configuration, both mechanisms generate antiphase oscillations (top panel). Physiological variability (see Methods for a description of variability) in both the synaptic (20\% variability in $\overline{g}_{syn}$) and cellular (20\% variability in $\overline{g}_{Ca,T}$) properties makes the oscillations unstable with Mechanism A but not with Mechanism B (bottom panel).
}
\label{fig:fig1}
\end{figure}

The transient depolarization following the termination of an hyperpolarizing input is an essential cellular property for the network rhythm, best known in the literature as post-inhibitory rebound (PIR)~\cite{Perkel1974}. Two major ionic currents have been shown to underlie the PIR (Figure~\ref{fig:fig1}A): i)~the hyperpolarization-activated cation current, $I_H$, an hyperpolarization-activated inward current that contributes to rebound responses in a diverse array of neurons in invertebrates and vertebrates~\cite{Angstadt2005}; ii)~the low-threshold T-type calcium current, $I_{Ca,T}$, which is deinactivated by hyperpolarization and then activates upon release from inhibition~\cite{Steriade1990}. Many studies have highlighted the distinction between these two currents in HCOs from an ``escape or release'' mechanism perspective, T-type calcium currents inducing the release mechanism and $I_H$-like currents promoting the escape mechanism~\cite{Wang1992, Skinner1994, Daun2009}): ): either the active cell ``releases'' its inhibitory effect on the silent cell (release mechanism), or the silent cell ``escapes'' from inhibition via the activation of an $I_H$-like current (escape mechanism).

Separately, the two types of current generate similar PIR traces in single cells (Figure~\ref{fig:fig1}A; see Methods for cellular models and model difference between Mechanism A and B). While both mechanisms are redundant for the generation of oscillations in a network with reciprocal inhibition, we emphasize a fundamental difference between the two: in presence of physiological variability, i.e., variability in the intrinsic cellular properties and synaptic connections (see Methods for a description of variability), only the rhythm generated by Mechanism B is robust (Figure~\ref{fig:fig1}B).This robustness property highlights a fundamental difference between the two mechanisms.

\begin{figure}[!h]
\begin{center}
\includegraphics[width=0.95\textwidth]{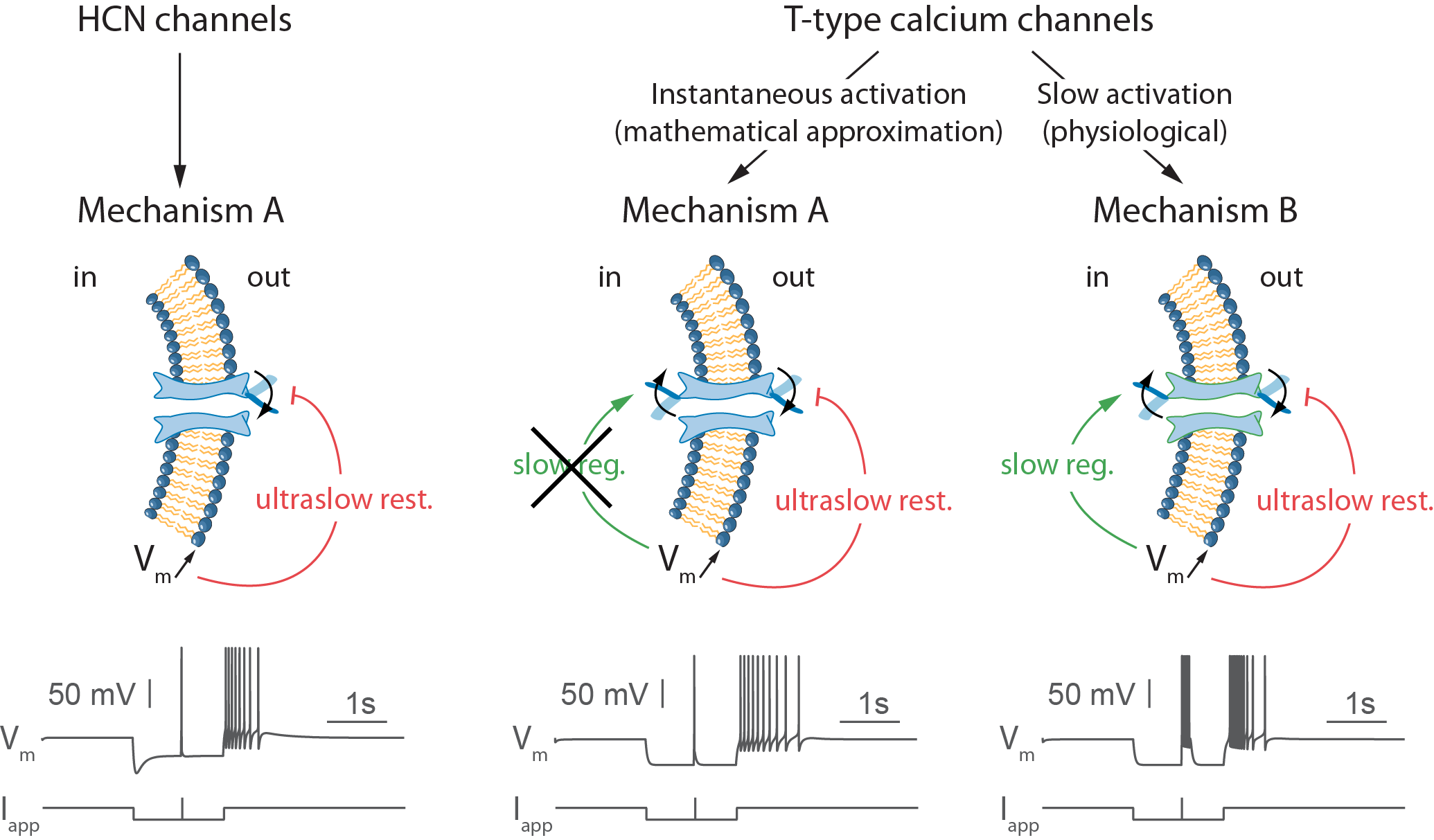}
\end{center}
\caption{
{\bf The slow activation of T-type calcium channels is the distinctive difference between the two PIR mechanisms.} Top: schemes representing ion channel gating in different cases. Bottom: responses of membrane potential ($V_{m}$) to a varying external applied current ($I_{app}$) for each case. Both mechanisms trigger a PIR via an ultraslow inward current in response to hyperpolarization, which brings ultraslow restorativity to the neuron (see Methods for cellular models). In addition, T-type calcium channels in Mechanism B, due to their slow activation, are a source of slow regenerativity. An instantaneous activation of the T-type calcium channels, \textit{i.e.}, steady-state approximation of their activation, suppresses this slow regenerativity and produces a Mechanism A PIR. Mechanism B PIR is endogenous as revealed by the specific signature during hyperpolarization.
}
\label{fig:fig2}
\end{figure}

This difference lies in the dynamical feedback loops generated by the gating variables of the two currents. Both currents generate an ultraslow (\textit{i.e.}, that lasts over the course of several action potentials) inward current in response to hyperpolarization, which is the foundation of the PIR. This inward current counteracts the external hyperpolarization and acts as a source of negative feedback on membrane potential variations---or restorativity in the terminology of ~\cite{Franci2013}---in the ultraslow timescale (Figure~\ref{fig:fig2}, red feedback loops). But, in contrast to $I_{H}$, $I_{Ca,T}$ provides a slow \textit{positive} feedback on membrane potential variations---or regenerativity in the terminology of~\cite{Franci2013}---via its slow activation variable (Figure~\ref{fig:fig2}, green feedback loop). This slow positive feedback is absent in Mechanism A. At the cellular level, the slow positive feedback is revealed by a specific signature during hyperpolarization~\cite{Franci2013}: a transient excitatory pulse that triggers a single spike in Mechanism A ($I_{H}$), triggers a burst in Mechanism B ($I_{Ca,T}$) (Figure~\ref{fig:fig2}, bottom panel). This signature reveals that bursts are endogenously generated with a PIR with slow regenerativity (Mechanism B), as opposed to a purely restorative---\textit{i.e.}, only $I_H$---PIR (Mechanism A).

A frequent modeling simplification is to neglect the slow activation kinetics of T-type calcium channels and to consider the activation at steady-state (\textit{i.e.}, instantaneous). It should be noted that the slow regenerativity is lost in this approximation, which eliminates the bursting signature observed in Mechanism B, as illustrated in Figure~\ref{fig:fig2}, center panel. 

In the rest of the paper, we investigate the impact of the difference between Mechanism A and B at the network level. More specifically, we look at a few simple quantities that characterize the network rhythm: the network---or interburst---frequency, \textit{i.e.}, the inverse of the time duration between two burst onsets, the duty cycle, which is the ratio between a burst duration and the time duration between two bursts, and the ratio between the duty cycle in neuron 1 and in neuron 2. In order to isolate the contribution of the slow regenerativity only, we compare the two mechanisms in a model that only includes, $I_{Na}$, $I_{K,DR}$, and one ``PIR current'', $I_{PIR}$, modeled by T-type calcium channels with instantaneous activation for Mechanism A, designated simply by ``PIR''f or the rest of the paper, and modeled by T-type calcium channels with slow activation for Mechanism B, designated by ``PIR + slow regenerativity'' (see Methods for cellular models and details of the simulations). Accordingly, the cellular models contain an identical PIR current in the two cases except for the activation time constant of the PIR current and both neuron models possess identical I/V curves. Results do not depend on other properties such as the role of the sag brought by $I_H$ or the difference between release and escape, both models differing only in their slow regenerativity. We stress that all results obtained under Mechanism A can be reproduced in a model where the PIR is modeled by $I_H$ channels only.

\subsection*{Robustness of network oscillations requires PIR with slow regenerativity}
There exists extensive experimental evidence that the rhythmic activity of neuronal circuits is robust against variability in intrinsic parameters (such as ionic conductances across neurons), extrinsic parameters (such as synaptic conductances), and exogenous noise (such as synaptic currents external to the circuit)~\cite{Liu1998, Golowasch1999, Golowasch1999a, Goldman2001, Destexhe2012}. We tested the robustness of HCOs in a network with PIR without slow regenerativity against a network with PIR with slow regenerativity (see Methods for network description and details of the simulations). The results show the drastic influence of cellular slow regenerativity in the robustness of the network.

\begin{figure}[!h]
\begin{center}
\includegraphics[width=0.95\textwidth]{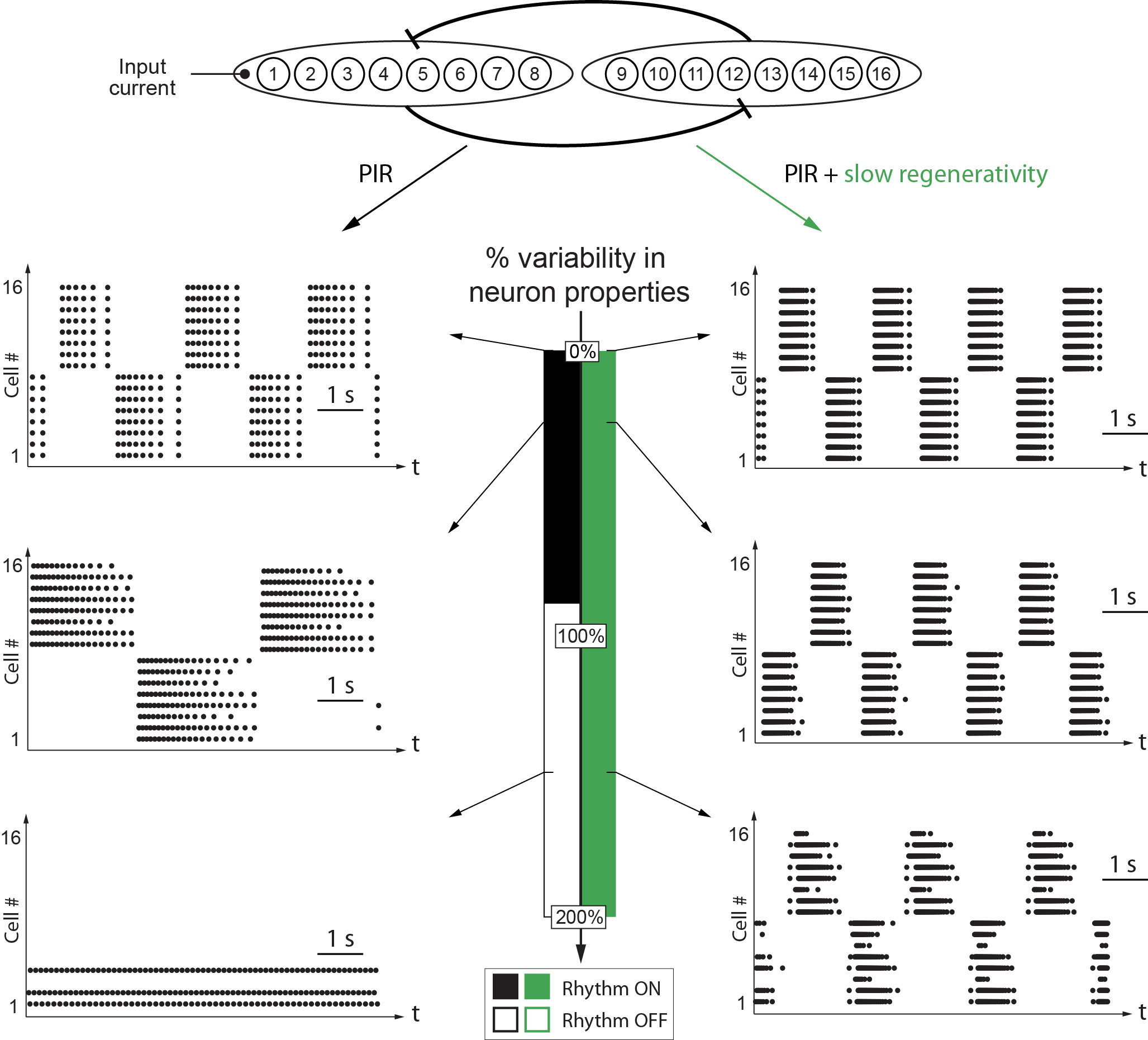}
\end{center}
\caption{
{\bf Slow regenerativity makes network oscillations insensitive to intrinsic variability.} Network oscillations are robust towards intrinsic variability only with slow regenerativity (see Methods for a description of variability). Left panel: PIR only. Right panel: PIR + slow regenerativity. Variability (level: 0\% to 200\%) in the maximal conductance of the PIR current, $\overline{g}_{PIR}$. Filled colors indicate presence of rhythmic activity and blanks indicate no rhythmic activity (see Methods for detection of rhythm). Raster plots with 0\%, 25\%, and 150\% variability, respectively. Network connections: all the neurons in the first population are connected to all the neurons in the second population, and vice versa. 
}
\label{fig:fig3}
\end{figure}

Intrinsic variability of the network was studied by introducing variability (see Methods for a description of variability) in the maximal conductance of the PIR current, $\overline{g}_{PIR}$, in a network with two populations (Figure~\ref{fig:fig3}, network connections). Variability in the cellular properties dramatically impacts the rhythmic activity of the network without slow regenerativity (Figure~\ref{fig:fig3}, left panel). The network rhythm becomes unstable beyond 75\% of variability and is significantly perturbed for smaller values. In sharp contrast, the network oscillations with slow regenerativity are robust against intrinsic variability up to 200\% (Figure~\ref{fig:fig3}, right panel). Remarkably, the network frequency is almost unaffected by the intrinsic variability, a consequence of the positive feedback brought by slow regenerative currents. Instead, the network frequency is strongly affected by intrinsic variability without slow-regenerative currents.

\begin{figure}[!h]
\begin{center}
\includegraphics[width=0.95\textwidth]{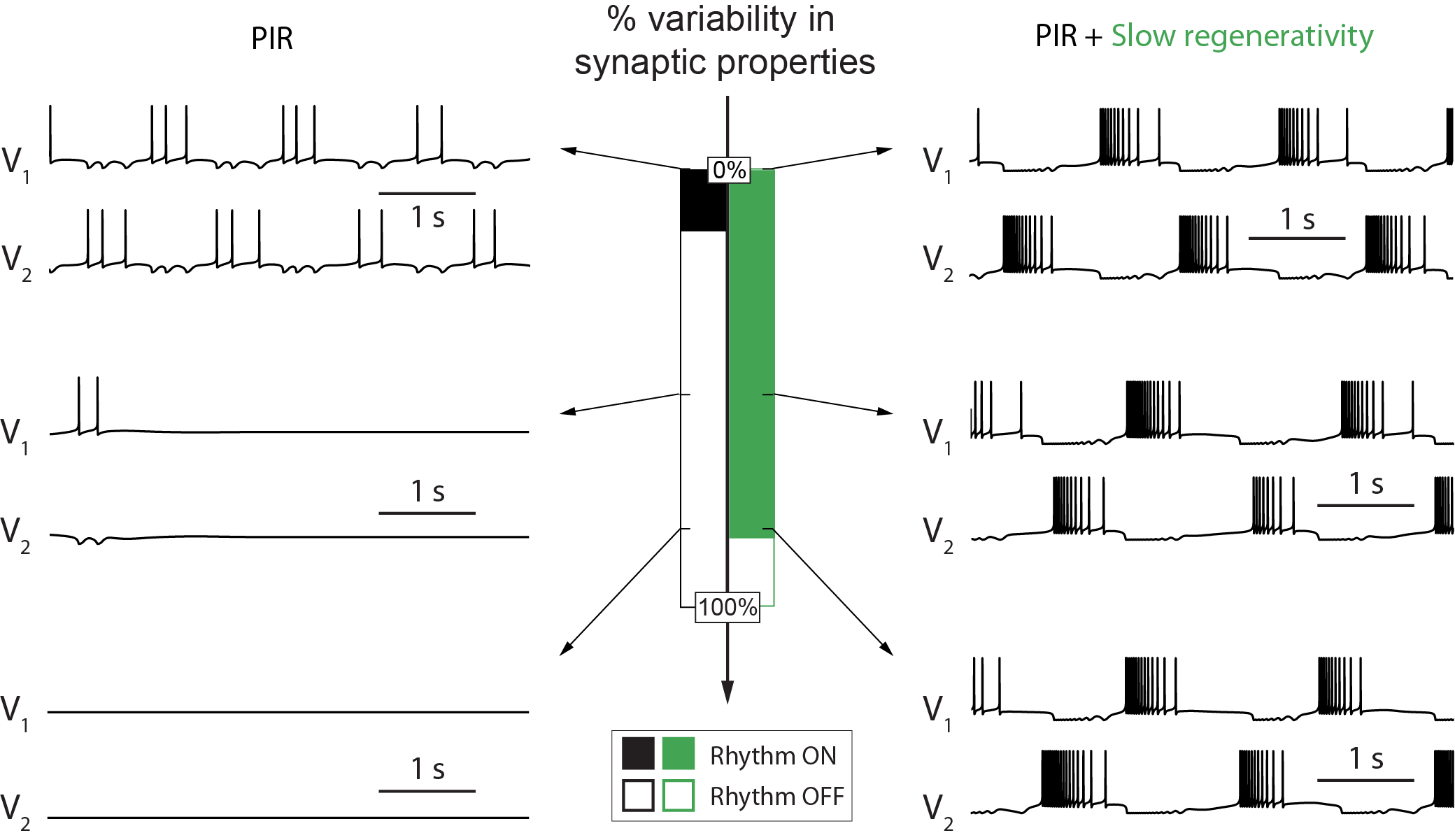}
\end{center}
\caption{
{\bf Slow regenerativity makes network oscillations insensitive to extrinsic variability.} Network oscillations are robust towards synaptic variability only with slow regenerativity (see Methods for a description of variability). Left panel: PIR only. Right panel: PIR + slow regenerativity. Variability (level: 0\% to 100\%) in the maximal conductance of the synaptic connection, $\overline{g}_{syn}$. Filled colors indicate presence of rhythmic activity and blanks indicate no rhythmic activity (see Methods for detection of rhythm). Membrane potentials with 0\%, 50\%, and 80\% variability, respectively.
}
\label{fig:fig4}
\end{figure}

The robustness of the network oscillations against variability in extrinsic parameters was studied by varying the maximal synaptic conductance parameters, $\overline{g}_{syn}$, in a two-neuron network with reciprocal connections (see Methods for a description of variability). Without slow regenerativity, a small variability in the synaptic conductances affects dramatically the network activity (Figure~\ref{fig:fig4}, left panel): identical maximal synaptic conductances generate oscillations but oscillations become unstable when the maximal synaptic conductances differ between the two cells. Oscillations with a PIR without slow regenerativity are fragile to network variability. In sharp contrast, variability in the synaptic conductances is possible for a much larger range with slow regenerativity and the network frequency is also almost independent of the synaptic variability (Figure~\ref{fig:fig4}, right panel). Oscillations persist up to a variability higher than 80\%. A source of slow-positive feedback in the PIR mechanism is therefore essential to robustness of network oscillations against network variability.

\begin{figure}[!h]
\begin{center}
\includegraphics[width=0.95\textwidth]{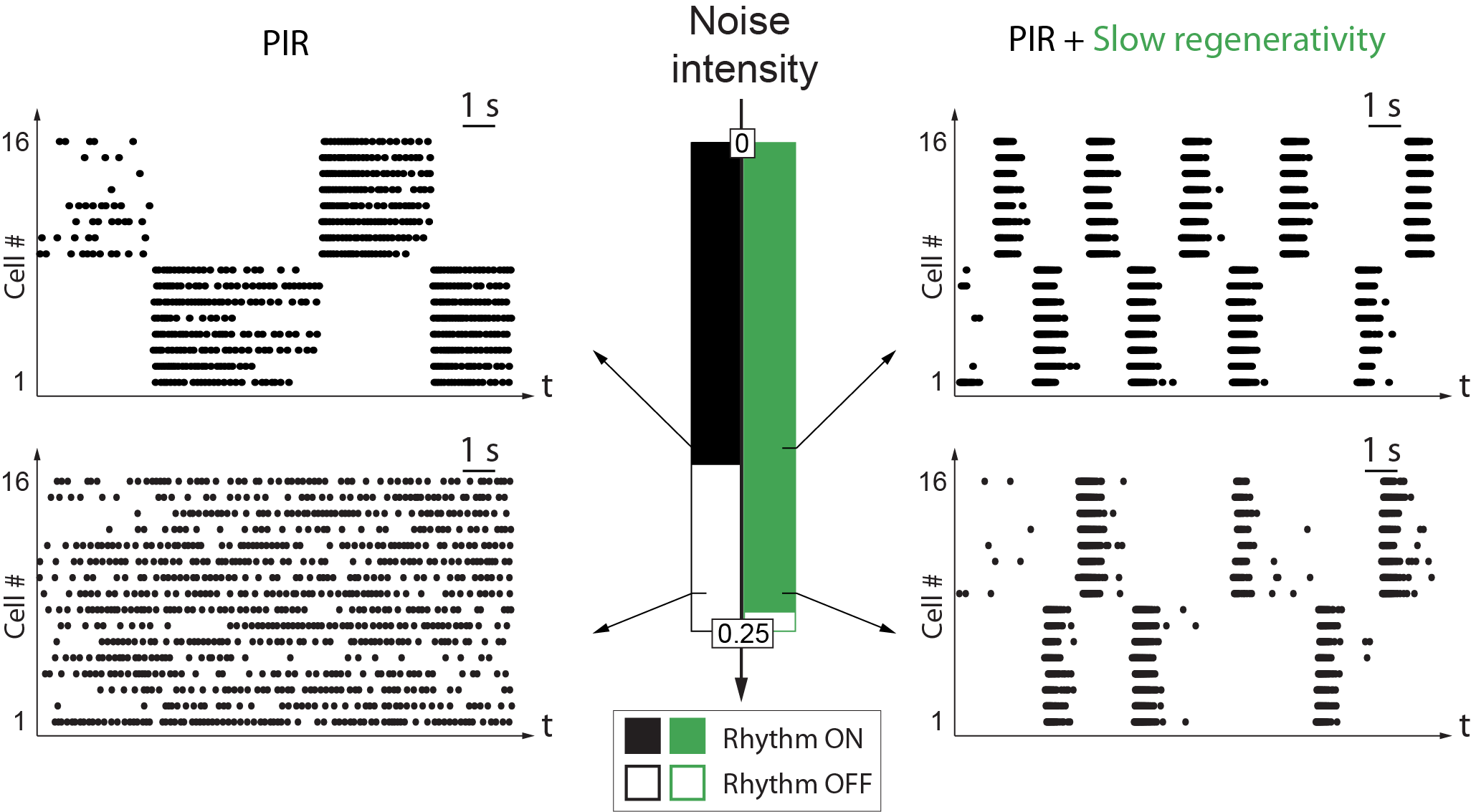}
\end{center}
\caption{
{\bf Slow regenerativity makes network oscillations robust against exogenous noise.} Network oscillations are robust towards exogenous noise only with slow regenerativity (see Methods for a description of variability). Left panel: PIR only. Right panel: PIR + slow regenerativity. Gaussian white noise (noise intensity $D$ ranges from 0 to 0.25 (in [$mV^2$])) is added to the neurons (see Methods for a description of noise). Filled colors indicate presence of rhythmic activity and blanks indicate no rhythmic activity (see Methods for detection of rhythm). Raster plots with noise intensity $D$ of 0.150$mV^2$ and 0.225$mV^2$, respectively.
}
\label{fig:fig5}
\end{figure}

The robustness of the network oscillations against exogenous disturbances was investigated by adding a Gaussian white noise in the equation that models membrane potential variations (see Methods for a description of noise). This emulates the external perturbations---spike train inputs from surrounding neurons---received by a network when studied in a noisy environment rather than in isolation~\cite{Lindner2003}. We simulated a sixteen-neuron network with two populations, with a different noise source for each neuron (Figure~\ref{fig:fig5}). 

The results are consistent with the robustness against parameter variability. Without slow regenerativity, oscillations are sensitive to noise and completely disappear with a noise level greater than 0.15$mV^2$. With slow regenerativity, oscillations are robust to noise up to a level of 0.225$mV^2$. Similarly to the introduction of variability in intrinsic and extrinsic parameters, the network frequency is also less affected with in the presence of slow regenerativity.

\subsection*{Robust modulation of network properties requires slow regenerativity}

Neuromodulators can tune and reconfigure the network dynamics, affecting both the frequency and phasing of neurons~\cite{Marder1992, Marder2002}. For instance, in the Tritonia swim CPG, intrinsic modulation can produce an enhanced level of excitability, triggering circuit activity that is maintained after the initial signal, generating an escape swimming response to particular aversive stimulus~\cite{Katz1994}. Neuromodulation can also switch the circuit between rhythms: in the crustacean stomatogastric ganglion, neuromodulators can switch the circuit activity from the fast pyloric rhythm, to two slower rhythms, the gastric mill rhythm and the cardiac sac rhythm~\cite{Marder2012}. In addition, neuromodulators can determine the active neuronal elements in the circuit or combine elements from different circuits into one~\cite{Marder2012, Harris-Warrick2011}.

Experimentally, the network properties, \textit{i.e.}, network frequency and duty cycle---or phase relation---can be modulated via both intrinsic neuron parameters and synaptic parameters on multiple timescales~\cite{Harris-Warrick1991, Fellous1998, El-Manira2000, Marder2002, Dickinson2006, Harris-Warrick2011, Marder2012, Dayan2012, Nadim2014}. In this section, we investigate how the network rhythmic activity responded to these modulations, both with PIR without slow regenerativity and with PIR with slow regenerativity (see Methods for network description and details of the simulations). 

Extrinsic parameters, \textit{i.e.}, the synaptic parameters $\overline{g}_{syn}$ and $\tau _{syn}$, given intrinsic (cellular) characteristics, modulate the network frequency. Synaptic coupling is very plastic~\cite{Gerstner2002, Dayan2005} and synapses are a primary target of modulators~\cite{Nadim2014}. Synaptic currents can be generated by the cooperation of several ion channel subtypes which can have slightly different kinetics. Variation of the synaptic parameters results from a variation of the contribution of all the subtypes. Absolute variation of the different ion channels influences the maximal conductance whereas their relative variation can modulate the time constant of the synaptic current that aggregates all the different subtypes in a model. Therefore, both the synaptic magnitude, $\overline{g}_{syn}$, and the synaptic kinetics, $\tau _{syn}$, can be sources of modulation.

\begin{figure}[!h]
\begin{center}
\includegraphics[width=0.95\textwidth]{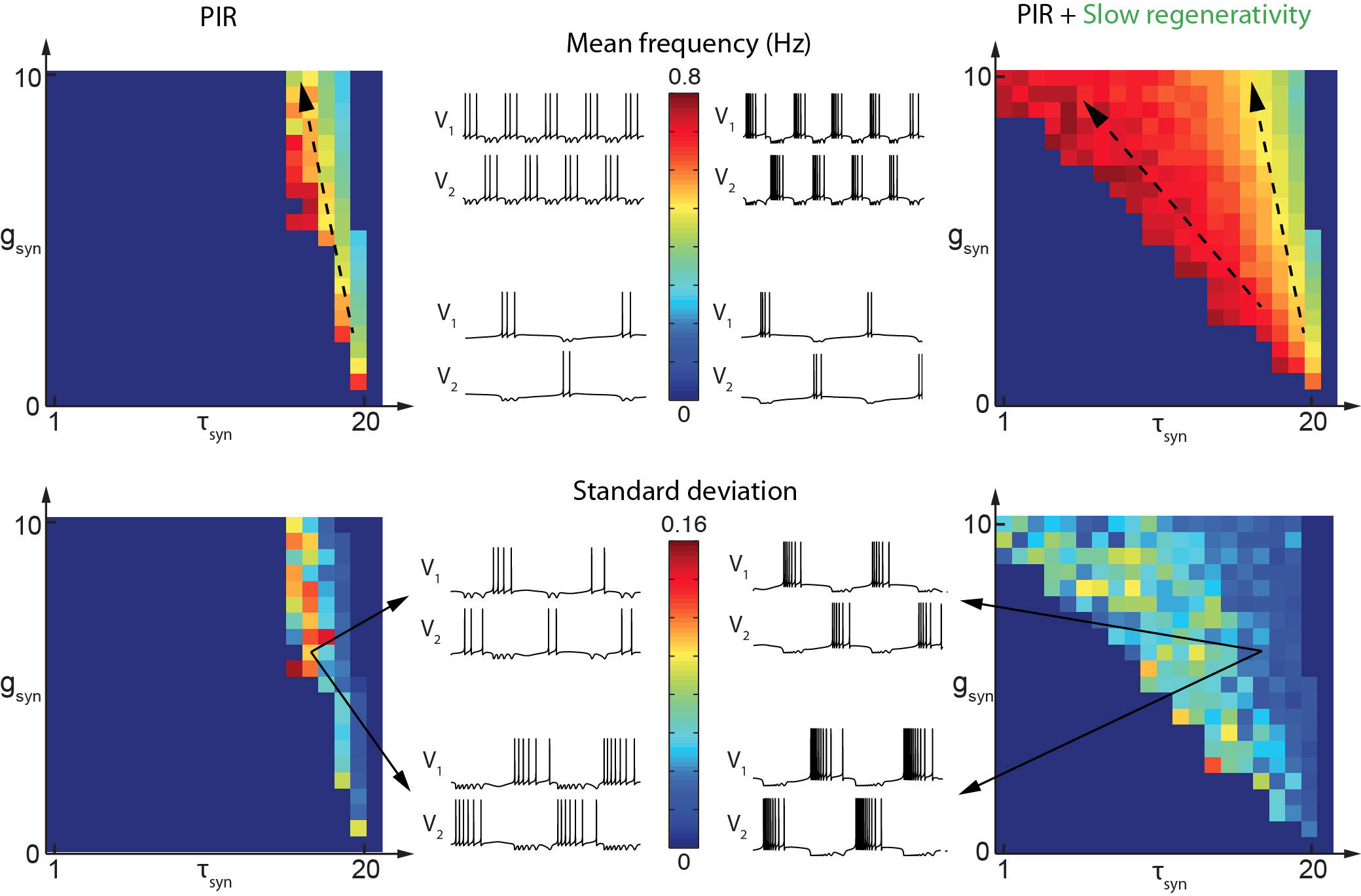}
\end{center}
\caption{
{\bf Frequency modulation with extrinsic parameters is fragile without slow regenerativity.} Modulation of the network frequency by varying synaptic parameters, $\overline{g}_{syn}$ (in [$mS/cm^2$]) and $\tau _{syn}$ (in [$ms$]), is robust with slow regenerativity but fragile without. Left panel: PIR only. Right panel: PIR + slow regenerativity. Mean frequency (top panel) and standard deviation (bottom panel) for ten simulations with 40\% variability in $\overline{g}_{syn}$ and 20\% variability in $\overline{g}_{PIR}$ (see Methods for a description of variability, mean frequency, and standard deviation). Membrane potentials top panel: maximal and minimal oscillation frequency, respectively. The arrows indicate the direction of frequency modulation. Membrane potentials bottom panel: two different simulations with the same $\overline{g}_{syn}$ and $\tau _{syn}$ parameters, $\overline{g}_{syn}$ and $\overline{g}_{PIR}$ are affected by parameter variability.
}
\label{fig:fig6}
\end{figure}

Oscillations with cellular slow regenerativity can be modulated over a large range by extrinsic parameters (Figure~\ref{fig:fig6}, right panal). Variation of the $\overline{g}_{syn}$ and $\tau _{syn}$ parameters generates a 150\% increase in network frequency (see Methods for a description of mean frequency). Such a span of modulation is observed in physiological rhythms: for instance, there is a 150\% increase in frequency from slow-wave sleep ($\approx 4 Hz $) to sleep spindles ($\approx 10 Hz $) and a 250\% increase from beta-band oscillations ($\approx 20 Hz $) to gamma-band oscillations ($\approx 70 Hz $). In addition, with slow regenerativity, the network frequency is only weakly sensitive to the variability in parameters as shown with the standard deviation plot and the highly similar network frequencies in the membrane voltage traces (see Methods for a description of variability and standard deviation). In opposition, variations of $\overline{g}_{syn}$ without slow regenerativity allow for network frequency modulation for a much smaller parameter range (Figure~\ref{fig:fig6}, left panel). Moreover, this modulation is very fragile and very sensitive to variability: the standard deviation reaches higher values than with slow regenerativity and membrane potential traces, for a same set of parameters but different simulations, are drastically different (Figure~\ref{fig:fig6}, bottom left panel). Variation of $\tau _{syn}$ is almost impossible: $\tau _{syn}$ must lie in a very specific timescale for the oscillations to develop in the network. The modulation requires a tight coupling between intrinsic and extrinsic parameters: the network oscillations are a direct reflection of the unicellular activity. The oscillation frequency is set by the neuron intrinsic dynamics and almost no variation can be induced by the synaptic dynamics.

Intrinsic parameters, \textit{i.e.}, the cellular parameters $\overline{g}_{PIR,1}$ and $\overline{g}_{PIR,2}$, given extrinsic (synaptic) characteristics, modulate the duty cycle and duty cycle ratio (see Methods for a description of duty cycle and duty cycle ratio). Many neuromodulators act on the neuron intrinsic properties by altering the balance of conductances, modifying their excitability properties~\cite{Marder2002}. The maximal conductance of the PIR current is a natural candidate for modulation by intrinsic parameters.

\begin{figure}[!h]
\begin{center}
\includegraphics[width=1\textwidth]{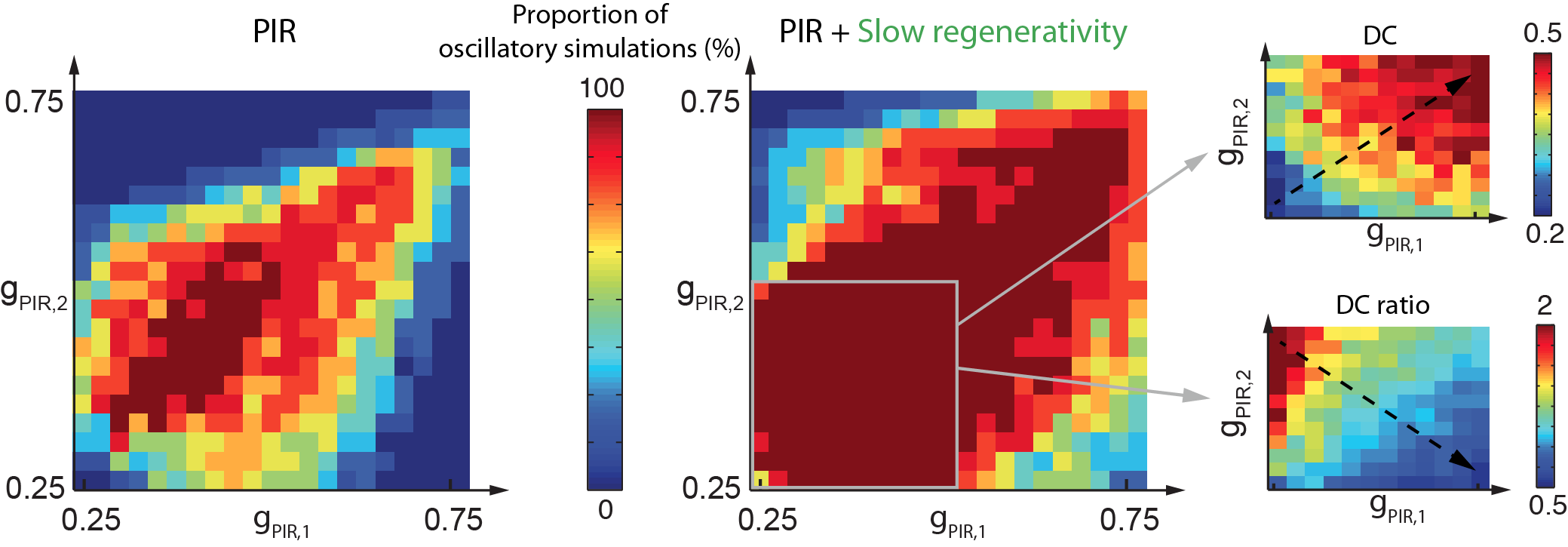}
\end{center}
\caption{
{\bf Duty cycle modulation with intrinsic parameters is fragile without slow regenerativity.} Modulation of the duty cycle and duty cycle ratio by varying intrinsic parameters of neuron 1, $\overline{g}_{PIR,1}$ (in [$mS/cm^2$]) and, neuron 2, $\overline{g}_{PIR,2}$ (in [$mS/cm^2$]), is robust with slow regenerativity but fragile without. Left panel: PIR only. Right panel: PIR + slow regenerativity. Proportion of simulations with stable rhythmic activity for ten simulations with 40\% variability in $\overline{g}_{syn}$ and 20\% variability in $\overline{g}_{PIR}$ (see Methods for a description of variability, detection of rhythm, and proportion of oscillatory HOCs). For the case with slow regenerativity, zoom in the stable region for mean duty cycle, (DC, top panel) and mean duty cycle ratio, (DC ratio, bottom panel) from the ten simulations (see Methods for a description of duty cycle and duty cycle ratio). The arrows indicate the direction of DC and DC ratio modulation. 
}
\label{fig:fig7}
\end{figure}

The high robustness brought by cellular slow regenerativity allows also for the modulation by intrinsic parameters even in presence of variability in the network (Figure~\ref{fig:fig7}; see Methods for a description of variability). Covariation of the maximal PIR conductances, $\overline{g}_{PIR,1}$ and $\overline{g}_{PIR,2}$, leads to an increase in duty cycle ratio of 150\% (Figure~\ref{fig:fig7}, top right panel). Independent variation of the same parameters, \textit{i.e.}, varying $\overline{g}_{PIR,1}$ and $\overline{g}_{PIR,2}$ independently, modulates the duty cycle ratio up to a factor two (Figure~\ref{fig:fig7}, bottom right panel). Variation in phase relation have been observed for instance in cats, during normal locomotion, where the shortening, by a factor two or three, of one of the phase (the extensor phase) leads to faster walking~\cite{Halbertsma1983}. In contrast, our computational model suggests that modulation with PIR without slow regenerativity is so fragile that it is unrealistic. Stable oscillations with variation of intrinsic parameters do not cover a large parameter range (Figure~\ref{fig:fig7}, left panel).

In brief, the high robustness brought by cellular slow regenerativity allows for the modulation by both extrinsic and intrinsic parameters. In addition, the network frequency and duty cycle can be modulated independently and in presence of physiological variability in the network.

\section*{Discussion}

\subsection*{Cellular slow regenerativity is essential to robustness and modulation of network rhythmic activity}

The main message of this paper is to highlight the role of {\it slow regenerativity}, a cellular excitability property, in endowing network oscillations with robustness and modulation properties that seem ubiquitous in physiological neuronal networks. An ionic current is slowly regenerative if it provides a source of positive feedback around resting potential in the slow timescale of repolarization~\cite{Franci2013}. The importance of this cellular property was assessed in one of the simplest and best understood network oscillation mechanisms, the antiphasic rhythm observed between two populations of neurons reciprocally connected by inhibitory synaptic connections. Many earlier studies have emphasized the role of post-inhibitory rebound (PIR) at the cellular level as a core mechanism for the network oscillation, and have identified $I_H$ and $I_{Ca,T}$ as two distinct ionic currents that can participate in the PIR. Our novel contribution is to observe that the cellular PIR will enable a robust and subject to modulation network oscillation only in the presence of a slow-regenerative ionic current. Because both $I_H$ and $I_{Ca,T}$ are sources of PIR currents but only $I_{Ca,T}$ is slow regenerative, our paper suggests a novel and somewhat fundamental complementarity between T-type calcium and $I_H$ channels in PIR mechanisms.

As a source of positive feedback, regenerative currents make the PIR endogenous, that is, robust to intrinsic and extrinsic sources of variability. As a consequence, a PIR with slow-regenerative currents allows for network oscillations that are robust and subject to modulation. The network oscillation is robust because it can sustain large variability across the neuronal population both in intrinsic (cellular) and extrinsic (synaptic) parameters. It is also subject to modulation because the frequency and phase properties of the oscillation can be controlled over a broad range by a relative modulation of extrinsic or intrinsic conductances. Our computational investigation illustrated that this robustness and modulation properties are lost when the PIR is purely ultraslow restorative.

\subsection*{Complementarity between the two types of PIR currents}

In the context of HCOs, many neurons possess both $I_H$ and $I_{Ca,T}$, the two main currents that contribute to PIR~\cite{Calabrese1992, DeSchutter1993, Nadim1995, Cymbalyuk2002, Sorensen2004, Olypher2006, Doloc-Mihu2011}. When those two currents are present, both $I_H$ and $I_{Ca,T}$ can be a source of modulation. In this case, the presence of T-type calcium currents as a source of slow regenerativity is sufficient to guarantee network oscillation robustness. On the other hand, the hyperpolarization-activated cation current can modulate drastically the network frequency and duty cycle~\cite{Cymbalyuk2002, Sorensen2004}. However, our computational model suggests that this is only the case if a slow-regenerative current, and therefore cellular endogenous characteristics, is supplied by another mechanism. It is noteworthy that the necessity for slow restorativity can be achieved by other means, such as the presence of high-threshold calcium channels. This necessary condition for slow regenerativity reveals a somewhat fundamental complementarity, distinct from the release or escape view, between the two channels: $I_{Ca,T}$ allows for stable rhythmic oscillations to emerge and $I_H$ enlarges the modulation possibilities.

\subsection*{Positive feedback as a source of endogenous activity}

Slow regenerativity is nothing but a source of positive feedback in the slow timescale of repolarization. It is a slow analog of the positive feedback brought by sodium activation in the fast timescale  of spike upstroke. In previous work~\cite{Franci2013}, we showed that this positive feedback is essential for the robust coexistence of hyperpolarized and spiking states at the cellular level. We subsequently showed in~\cite{Franci2014} that this positive feedback is essential for modulation and robustness of bursting activities. Here we show that the same positive feedback at the cellular level is also essential for robustness and modulation at the network level. The common feature of the positive feedback in those three phenomena is that it makes the neuronal excitability in the slow timescale an endogenous property, robust to intrinsic and extrinsic variability. Making an activity endogenous is the very nature of positive feedback and has been emphasized in a number of contexts. The results presented in this paper are in line with the discussion of the role of positive feedback in other biological models, such as for instance the biochemical mechanisms underlying the mitotic oscillator~\cite{Novak1993, Pomerening2003, Tyson2008a}: the oscillator is endogenous and robust in the presence of positive feedback, whereas it becomes exogenous and entrainable when the source of positive feedback disappears. The importance at the network level of positive feedback at the cellular level is thought to be general and not specific to the case study of HCOs chosen in this paper for its simplicity and physiological relevance.

\subsection*{Slow regenerativity in half-center oscillator models}

There is a rich literature on computational models of oscillations generated by reciprocal inhibition. HCOs have been used to model rhythmic motor outputs in many invertebrates and vertebrates~\cite{Brown1911, Perkel1974, Satterlie1985, Calabrese1992, Marder1996, Harris-Warrick2011}. In a different context, models of thalamocortical spindle oscillations suggest that the rhythm originates from the thalamic reticular nucleus, which consists in interacting inhibitory nonoscillatory neurons~\cite{Wang1992, Wang1993, Destexhe1994, Golomb1994}.

It is of interest to observe the varying degree of cellular regenerativity in published models of HCOs. Early models are conductance-based and usually include at the cellular level both $I_H$ and $I_{Ca,T}$, the two main physiological currents eliciting the PIR~\cite{Huguenard1992a, Calabrese1992, DeSchutter1993, Nadim1995, Destexhe1994}. However, network computational studies often lead to a subsequent mathematical simplifications of the cellular details and the cellular slow-positive feedback is often lost in this reduction process. A frequent simplification in the literature (see e.g.~\cite{Wang1992, Wang1993, Golomb1993, Rinzel1998, Daun2009}) is to resort to a steady-state approximation of the calcium activation in the same way as it is normally done for sodium activation. But this approximation rests on neglecting {\it fast} dynamics, which amounts to consider calcium channels as a source of {\it fast} rather than {\it slow} positive feedback. The resulting reduced models have therefore lost their source of slow regenerativity, which makes them unsuitable for robustness and modulation studies at the network level. 

The alternative model reduction consists to model the cellular level as Morris-Lecar type of neurons, retaining the slow calcium currents but neglecting the fast sodium currents~\cite{Skinner1993, Skinner1994}. Those models do retain the slow-positive feedback source necessary for robustness but they lose the modulation capabilities illustrated in the present paper because the network interconnection properties are spike-dependent. This prevents exogenous modulation of the rhythm. In addition, if sodium spikes were added to a Morris-Lecar neuron with the addition of the spike currents (as suggested in~\cite{Kopell2002}) while keeping the calcium activation at steady-state, the slow regenerativity would be destroyed.

It should be highlighted that it is possible to derive reduced neuronal models that do retain the balance of slow positive and negative feedbacks as an explicit parameter, see \textit{e.g.} the recent models in~\cite{Drion2012, Franci2014}. The results of the present paper suggest that it is an important feature to retain in a simplified model aimed at network computational studies.

\section*{Methods}

All the numerical simulations and analyses were performed with MATLAB, MathWorks. The models were implemented in a MATLAB code and simulated using a forward Euler method with a time step of $0.005 ms$.

\subsection*{Cellular model} 
The cellular model is inspired from the crab stomatogastric ganglion (STG) conductance-based neuron model~\cite{Turrigiano1995, Liu1998}. The model contains the standard Hodgkin-Huxley (HH) currents (Hodgkin and Huxley, 1952): the transient sodium current, $I_{Na}$, a fast depolarizing current, and the delayed-rectifier potassium current, $I_{K,DR}$, a slower hyperpolarizing current, plus a leak current, $I_{L}$. The two currents responsible for PIR are: a low threshold T-type calcium current, $I_{Ca,T}$, and a hyperpolarization activated cation current, $I_H$. The membrane potential dynamics writes as follows:

\begin{subequations}
  \begin{align*}
  C \dot{V_m} = -I_{Na} - I_{K,DR} - I_{L} - I_{Ca,T} - I_{H} + I_{app},
  \end{align*}
\end{subequations}

where $C=1pF/cm^2$ is the membrane capacitance and $I_{app}$ is the applied current.

Each ionic current $i$ takes the standard HH form:

\begin{subequations}
  \begin{align*}
  I_i = \overline{g}_i m^{p_i} h^{q_i} \left(V_m-E_i \right),
  \end{align*}
\end{subequations}

where $\overline{g}_i$ is the maximal conductance for current $i$, $p_i$ and $q_i$ are integers, and $E_i$ is the reversal potential of the ion $i$. Table~\ref{tab:tab1} lists the values of $g$, $p$, $q$, and $E$ for the different currents.

\begin{table}[!ht]
\caption{
\bf{Parameters for the membrane currents of the cellular model.}}
\begin{tabular}{| c | c | c | c | c |}
\hline
& $g$ & $p$ & $q$ & $E$ \\
\hline
$I_{Na}$ & 60 & 3 & 1 & 50 \\
$I_{K,DR}$ & 40 & 4 & 0 & -70 \\
$I_{L}$ & 0.035 & 0 & 0 & -49 \\
$I_{Ca,T}$ & 0.3 & 3 & 1 & 120 \\
$I_{H}$ & 0.04 & 1 & 0 & -20 \\
\hline
\end{tabular}
\begin{flushleft}Notation is explained in the text. All conductances are in $mS/cm^2$ and membrane potentials in $mV$.
\end{flushleft}
\label{tab:tab1}
 \end{table}

Activation and inactivation variable dynamics follow the classical formalism:

\begin{subequations}
  \begin{align*}
  \tau _m \dot{m} &= m_{\infty} - m,\\
   \tau _h \dot{h} &= h_{\infty} - h,
  \end{align*}
\end{subequations}

where the functions for $\tau _m$, $m_{\infty}$, $\tau _h$, and $h_{\infty}$ are given in Table~\ref{tab:tab2}. Note that, if $p_i=0$ and/or $q_i=0$, $\tau _m$ and $m_{\infty}$ and/or $\tau _h$ and $h_{\infty}$ are not listed, respectively.

\begin{table}[!ht]
\caption{
\bf{Functions for the membrane currents of the cellular model.}}
\begin{tabular}{| c | c | c | c | c |}
\hline
 & $m_{\infty}$ & $h_{\infty}$ & $\tau _m$ & $\tau _h$ \\
\hline
$I_{Na}$ & $\frac{1}{1+exp\left(\frac{V_m+35.5}{-5.29} \right)}$ & $\frac{1}{1+exp\left(\frac{V_m+48.9}{5.18} \right)}$ & $1.32-\frac{1.26}{1+exp\left(\frac{V_m+120}{-25.0} \right)}$ & $\frac{0.67}{1+exp\left(\frac{V_m+62.9}{-10.0} \right)} * \left(1.5+\frac{1}{1+exp\left(\frac{V_m+34.9}{3.6} \right)} \right)$ \\
$I_{K,DR}$ & $\frac{1}{1+exp\left(\frac{V_m+12.3}{-11.8} \right)}$ & & $7.2-\frac{6.4}{1+exp\left(\frac{V_m+28.3}{-19.2} \right)}$ & \\
$I_{Ca,T}$ & $\frac{1}{1+exp\left(\frac{V_m+57.1}{-7.2} \right)}$ & $\frac{1}{1+exp\left(\frac{V_m+82.1}{5.5} \right)}$ & $21.7-\frac{21.3}{1+exp\left(\frac{V_m+68.1}{-20.5} \right)}$ & $840-\frac{718.4}{1+exp\left(\frac{V_m+55}{-16.9} \right)} $ \\
 $I_{H}$ & $\frac{1}{1+exp\left(\frac{V_m+80}{6} \right)}$ & & $272+\frac{1499}{1+exp\left(\frac{V_m+42.2}{-8.73} \right)}$ & \\
\hline
\end{tabular}
\begin{flushleft}Notation is explained in the text. All time constants are in $ms$.
\end{flushleft}
\label{tab:tab2}
 \end{table}
 
In Figures~\ref{fig:fig3}-\ref{fig:fig7}, the only current responsible for the PIR is $I_{Ca,T}$, either with instantaneous activation, \textit{i.e.}, $m (t) \equiv m_{\infty} (V_m (t))$or with slow activation, \textit{i.e.}, $\tau _m$ of Table~\ref{tab:tab2} ($\overline{g}_{H}=0mS/cm^2$). The former case corresponds to Mechanism A or ``PIR'', whereas the latter entails Mechanism B or ``PIR + slow regenerativity'' (see text for details). In both cases, we denote this current and its maximal conductance as$I_{PIR}$ and $\overline{g}_{PIR}$. 
 
\subsection*{Network model} 

The inhibitory synaptic connections are gamma-aminobutyric acid (GABA) and are made with exponential synapses of the GABA$_{\textrm{A}}$ type. The synaptic current between two neurons takes the form ~\cite{Golomb1994}:

\begin{align}
I_{syn} &= \overline{g}_{syn}\left(V-V_{syn}\right) \frac{1}{N} \sum^{N}_{j=1} s_{Aj}, \\ \notag
\frac{ds_{Aj}}{dt} &= k_{fA} x_{\infty}(V_j) \left(1-s_{Aj}\right) - k_{rA} s_{Aj}, \\ \notag
x_{\infty}(V) &= \left[1+exp \left( - \left(V - \Theta _s \right) / \sigma _s \right) \right]^{-1} , \notag
\end{align}

where $V_{j}$ is the presynaptic membrane potential and $N$ the number of presynaptic neurons. If not stated otherwise, $\overline{g}_{syn}=4mS/cm^2$, $V_{syn}=-75mV$, $k_{fA}=2ms^{-1}$, and $k_{rA}=0.1ms^{-1}$, $\Theta _s=-45mV$, $\sigma _s=2mV$.

\subsection*{Variability} 

Physiological variability is modeled by randomly selecting the values for the parameter subjected to variability in an interval, called \emph{variation range}, centered on a given parameter value. The \emph{variability level} quantifies the width of this interval, in percentage of the given value. For instance, a 200\% variability in $\overline{g}_{PIR}$ means that, for a given parameter value of $0.3 mS/cm^2$, the variation range has a width of $200\% * 0.3 =0.6$ and is centered on $0.3$. The parameters are therefore randomly selected out of the interval $[0 mS/cm^2, 0.6 mS/cm^2 ]$.

\subsection*{Analyses} 

A network is categorized as having a stable rhythmic activity (rhythm ON) if all the neurons in the network are still bursting in the stationary state. Due to the specific network structure---connection all-to-all from one population to the other, and \textit{vice versa}---all the neurons in one population receive the same input (coming from all the neurons in the other population). Therefore, if all the neurons are bursting, the bursts have been elicited by the same transient hyperpolarizing input---bursting cannot happen without this hyperpolarization---and the bursts are synchronous, \textit{i.e.}, all the bursts overlap but not necessarily the spikes. This provokes HCO antiphasic oscillations. Bursting in neurons is detected by having two consecutive spikes less than $200ms$ appart. Practically, we detect busting in all neurons after the transient phase: due to the time constants in play, analyzing the last $3s$ of data was sufficient to have bursting in the two populations in the stationary state. If at least one neuron was not bursting during this time period, the network is categorized as having no stable rhythmic activity (rhythm OFF). 

The frequency is the inverse of the time duration between the beginning of two bursts, or period, averaged over the two neurons. The duty cycle is the ratio between the burst duration and the period, averaged over the two neurons. The duty cycle ratio is the ratio between the duty cycle in neuron 1 and in neuron 2. Each value for the mean frequency, mean duty cycle and mean duty cycle ratio was computed from 10 simulations with the same set of parameters but with 40\% variability in $\overline{g}_{syn}$ and 20\% variability in $\overline{g}_{PIR}$. Only simulations endowed with rhythmic activity in the sense defined previously were considered to compute the means and standard deviations. If no rhythmic activity was detected, the computed means and standard deviations were set to 0. The proportion of oscillatory HCO quantifies the percentage of simulations that showed rhythmic activity out of the 10 simulations.

\subsection*{Simulation details} 

Cell simulations of Figure~\ref{fig:fig1}, are performed using the cellular model described above, with $\overline{g}_{Ca,T}=0mS/cm^2$ for Mechanism A and $\overline{g}_{H}=0mS/cm^2$ for Mechanism B. The applied current, $I_{app}$, on both neurons takes a value of $-0.55nA$. During the hyperpolarization, the applied current drops to $-1.95nA$.

The network simulations of Figure~\ref{fig:fig1} were obtained with the cellular model with $\overline{g}_{H}=0mS/cm^2$, and with instantaneous activation for the left panel, \textit{i.e.}, $m_{Ca,T} (t) \equiv m_{\infty _{Ca,T}} (V_m (t))$ from Table~\ref{tab:tab2}, and slow activation with time constant $\tau _{m_{Ca,T}}$ (see Table~\ref{tab:tab2}) for the right panel. The top panel does not present any parameter variability, the two neurons and synaptic connections are identical. Physiological variability is simulated with 40\% variability in $\overline{g}_{syn}$ and 20\% variability in $\overline{g}_{Ca,T}$.

The cell models in Figure~\ref{fig:fig2} were performed using the cellular model described above, with $\overline{g}_{Ca,T}=0mS/cm^2$ in the left panel, $\overline{g}_{H}=0mS/cm^2$ and instantaneous activation in the center panel, and $\overline{g}_{H}=0mS/cm^2$ and slow activation in the right panel. The applied current, $I_{app}$, takes a value of $-0.55nA$. During the hyperpolarization, the applied current drops to $-1.95nA$ and takes a value of $10nA$ for $10ms$ for the fast depolarizing input.

The two populations of Figure~\ref{fig:fig3}, each composed of 8 neurons, are connected all-to-all with the synaptic current described in the previous section. The intrinsic neuron parameter is $\overline{g}_{PIR}=0.3mS/cm^2$ with a variability level from 0\% to a 200\%.

The simulations of Figure~\ref{fig:fig4} were done for a two-neuron network. The extrinsic neuron parameter is $\overline{g}_{syn}=4mS/cm^2$ with a variability level from 0\% to a 100\%.

The two populations considered in Figure~\ref{fig:fig5}, each composed of 8 neurons, are connected all-to-all with the synaptic current described in the previous section. A Gaussian white noise is added in the voltage equation to model the typical spike train input received from the many other unmodeled neurons~\cite{Lindner2003}. The noise is modeled by $\sqrt{2D} \xi(t)$, where $D$ is the noise intensity and varies from 0 to 0.25, and $\xi(t)$ is drawn from a normal distribution with zero mean and unitary standard deviation and is different for each neuron.

In Figure~\ref{fig:fig6}, the maximal synaptic conductance, $\overline{g}_{syn}$ (from $0mS/cm^2$ to $10mS/cm^2$), and the synaptic time constant, proportional to $1/k_{rA}$ ($k_{rA}$ varies from $0 ms^{-1}$ to $1 ms^{-1}$), vary simultaneously for the two neurons. The variability level is 40\% for $\overline{g}_{syn}$ and 20\% for $\overline{g}_{PIR}$. Membrane potential plots ($\overline{g}_{syn}$[$mS/cm^2$], $k_{rA}$[$ms^{-1}$]): top left panel--(5, 0.05) and (0.5, 0.05); top right panel--(5, 0.05) and (6.5, 0.5); bottom panel--(6, 0.2) with variability.

In Figure~\ref{fig:fig7}, the maximal PIR conductances, $\overline{g}_{PIR,1}$ and $\overline{g}_{PIR,2}$ (from $0.25mS/cm^2$ to $0.75mS/cm^2$), vary independently for the two neurons. The variability level is 40\% for $\overline{g}_{syn}$ and 20\% for $\overline{g}_{PIR}$. The zoom for the duty cycle and duty cycle ratio covers a range from $0.25mS/cm^2$ to $0.45mS/cm^2$.

\bibliography{RP_bilbi}

\section*{Figure Legends}

\end{document}